\documentclass[aps,twocolumn,amsmath,nofootinbib,floatfix]{revtex4}
\usepackage{graphicx}
\usepackage{latexsym}
\usepackage{dcolumn}
\usepackage{bm}
\tighten

\begin{document}
\title{Signal of the pion string at high-energy collisions}
\author{Tao Huang$^1$, Yunde Li$^{1,2}$, Hong Mao$^{1,4}$, Michiyasu Nagasawa$^{1,3}$ and Xinmin Zhang$^1$}
\address{1. Institute of High Energy Physics, Chinese Academy of Sciences, Beijing
100049, China  \\
2. Physics Department, East China Normal University
Shanghai, 200062, China\\
3.Department of Information Science, Faculty of Science, Kanagawa
University, Kanagawa 259-1293, Japan\\
4.Graduate School of the Chinese Academy of Sciences, Beijing
100039, China}

\begin{abstract}
We study the possible signals of a pion string associated with the
QCD chiral phase transition in LHC Pb - Pb collision at energy $
\sqrt{s}=5.5$ TeV. In terms of the Kibble-Zurek mechanism we
discuss the production and evolution of the pion string. The pion
string is not topologically stable, it decays into neutral pions
and sigma mesons which in turn decay into pions. Our results show
that all the neutral pions from the pion string are distributed at
the low momentum and the ratio of neutral to charged pions from
the pion string violates the isospin symmetry. For the momentum
spectra of the total pions, the signal from the sigma particle
decay which is from the pion string will be affected by the large
decay width of the sigma significantly.
\newline PACS number(s): 11.27.+d, 25.75.-q, 98.80Cq
\end{abstract}

\maketitle

\section{Introduction}
The formation of topological defects in phase transition is a very
generic phenomenon in physics. It can be studied experimentally in
different condensed matter systems. It is generally believed that
the early evolution of Universe undergoes a sequence of phase
transitions, and the produced topological defects in these phase
transitions may have observable consequences to the properties of
the universe today. For example, the cosmic strings have been
suggested as one possible source for the primordial density
perturbations that give rise to the large-scale structure of the
universe and the temperature fluctuations of the cosmic microwave
background (CMB) radiation\cite{Rajantie03}\cite{Villenkin00}. In
particular, in Ref.\cite{Brandenberger99} the effect of the pion
string on the primordial magnetic field generation in the early
universe has been considered and its cosmological significance is
pointed out. However, in this paper, we will turn from cosmology
to laboratory experiments and attempt to study the possible
signals of the pion string in the heavy ion collision experiments
which have many similarities with the nonequilibrium phenomena
that also take place in heavy ion collisions
experiments\cite{Rajagopal93}\cite{Rajagopal95}.

It is difficult to make experimental tests of our ideas about the
formation and evolution of topological defects in cosmology
directly. What we can do is to look for analogous processes in
experimentally accessible condensed-matter systems. Fortunately,
topological defects are formed at phase transitions in certain
condensed matter systems such as super-fluids and superconductors,
this phenomenon is theoretically very similar to its cosmological
counterpart, and we can use this analogy to do "cosmology
experiments"\cite{Zurek85}. On a more fundamental level, these
same experiments can be used to test our understanding of
non-equilibrium dynamics of quantum field
theories\cite{Volovik03}.

In relativistic nucleus-nucleus collisions, some phenomena which
like that happens in the Big Bang have been observed, this is
called the Little Bang. Thus as hadron momentum spectra
correlations provide strong evidence for the existence of the
Little Bang: thermal hadron radiation with T=90-100MeV and strong
3-dimensional (Hubble like) expansion with transverse flow
velocities 0.5-0.55c \cite{Heinz99}. So the study of Little Bang
from relativistic nucleus-nucleus collisions may construct a
bridge between high energy particle physics and the cosmology. On
the other hand, we are not yet in a position to give an evidence
that quark-gluon plasma(QGP) has really been produced. It turns
out to be a difficult task to figure out the theoretical picture
of QGP even in equilibrium. Beyond the trivial level of the trees
and the non-equilibrium properties of the QGP are essentially
unknown. There is no unique signal of QGP in the understanding of
nucleus-nucleus collisions so far. As pointed out by Rajantie
\cite {Rajan02}, since the heavy-ion collisions experiments are so
complicated that the reliable and accurate theoretical
calculations are needed in order to confront the experimental
results, but our present understanding of the theory is too
rudimentary for that.

Then the insight provided by condensed systems experiments is
therefore likely to be extremely useful. In particular, it is
believed that at a certain value of the beam energy, the QGP which
is produced in the collision cools through a second-order
transition point. Pion strings as well as other topological and
non-topological strings are expected to be
produced\cite{Zhang98}\cite{Balachandran02}. An early study on the
effects of these strings in the case of heavy ion collisions and
in the early universe has been performed in
Ref.\cite{Balachandran02}, in their paper, they speculate that
formation and subsequent evolution of the network of these string
defects can give rise to inhomogeneous distribution of baryons and
also the energy density by using the Kibble mechanism. In this
paper, we extend their works on the formation and evolution of
strings and discuss the possible signal of pion strings during
chiral phase transition in the high ion collisions.

So far the theoretical scenario what can be applied to study the
formation of topological defects in systems with global symmetry
is the Kibble-Zurek mechanism\cite{Zurek85}\cite{Kibble76}.
Moreover, according to Pisarski and Wilczek \cite{Pisarski84} the
chiral phase transition is expected to be of the second order for
two massless flavors, it is customary then in this paper to apply
the Kibble-Zurek mechanism in order to study the formation and
evolution of the pion string during Pb-Pb central collisions at
the LHC with energy $\sqrt{s}=5.5 TeV$.

The remainder of this paper is organized as follows. We give a
brief review of the pion string in the linear sigma model in
Sec.II. It is shown how to use the Kibble-Zurek mechanism to
consider the formation of the pion string at LHC in Sec.III. We
discuss the evolution and decay of the pion string and their
possible observational consequences at LHC in Sec. IV. Sec. V is
reserved for summary and discussion.

\section{The pion string in QCD}
The linear sigma model which serves as a good low energy effective
theory of the QCD was first introduced in the 1960s as a model for
pion-nucleon interactions\cite{Gell-Mann60} and has attracted much
attention recently, especially in studies involving disoriented
chiral condensates\cite{Rajagopal93}\cite{Kuraev03}. This model is
very well suited to describing the physics of pions in studies of
chiral symmetry. In what follows we will review the work of
\cite{Zhang98} in which it was shown that below the chiral
symmetry breaking scale, the linear sigma model admits global
vortex line solutions, the pion string. We then attempt to use the
Kibble-Zurek mechanism to give out a quantitative description of
the formation and evolution of the pion string and their possible
observational consequences in heavy ion collisions.

Considering a simple case of QCD with two massless quarks $u$ and
$d$, the Lagrangian of strong interaction is invariant under
$SU(2)_L\times SU(2)_R$ chiral transformation
\begin{equation}
\Psi _{L,R}\rightarrow \exp (-i\vec{\theta}_{L,R}\cdot \vec{\tau
})\Psi _{L,R},
\end{equation}
where $\Psi^{T} _{L,R}=(u,d)_{L,R}$. However this chiral symmetry
does not appear in the low energy particle spectrum since it is
spontaneously broken to the diagonal subgroup formation.
Consequently, three Goldstone bosons, the pions, appear and the
(constituent) quarks become massive. At low energy, the
spontaneous breaking of chiral symmetry can be described by an
effective theory, the linear sigma model, which involves the
massless pions $\vec{\pi }$ and a massive $\sigma $ particle. As
usual, we introduce the field
\begin{equation}
\Phi =\sigma \frac{\tau ^0}2+i\vec{\pi }\cdot \frac{%
\vec{\tau }}2,
\end{equation}
where $\tau ^0$ is the unity matrix and $\vec{\tau }$ is the Pauli
matrices with the normalization condition $\textrm{Tr}(\tau ^a\tau
^b)=2\delta ^{ab}$. Under $SU(2)_L\times SU(2)_R$ chiral
transformations, $\Phi$ transforms as
\begin{equation}
\Phi \rightarrow L^{+}\Phi R.
\end{equation}

The renormalizable effective Lagrangian of the linear sigma model
can be written as
\begin{equation}
{\mathcal{L}}={\mathcal{L}}_\Phi +{\mathcal{L}}_q,
\end{equation}
where
\begin{equation}\label{eq1}
{\mathcal{L}}_\Phi =\textrm{Tr}[(\partial _\mu \Phi )^{+}(\partial
^\mu \Phi )]-\lambda [\textrm{Tr}(\Phi ^{+}\Phi )-\frac{f_\pi^2
}{2}]^2
\end{equation}
and
\begin{equation}
{\mathcal{L}}_q=\overline{\Psi }_Li\gamma ^\mu \partial _\mu \Psi _L+\overline{\Psi }%
_Ri\gamma ^\mu \partial _\mu \Psi _R-2g\overline{\Psi }_L\Phi \Psi
_R+h.c..
\end{equation}
During chiral symmetry breaking, the $\sigma$ field takes on a
non-vanishing vacuum expectation value, which breaks $SU(2)_L\times SU(2)_R$%
~down to $SU(2)_{L+R}$. It results in a massive sigma particle
$\sigma $ and three
massless Goldstone bosons $\vec{\pi }$, as well as giving a mass $%
m_q=gf_\pi $ to the constituent quarks\footnote{The $\sigma$ field
can be used to represent the quark condensate and the order
parameter for the chiral phase transition since both exhibit the
same behavior under chiral transformations\cite{Birse94}, the
pions are very light particles and can be considered approximately
as massless Goldstone bosons.}.

In the previous paper, two of us (X.Z. and T.H.) with
Brandenberger\cite{Zhang98} discovered a type of classical
solution, the pion string, in the above linear sigma model. This
pion string is very much like the spin vortex produced at the
superfluid transition of the liquid $^3He$\cite{Volovik03}.
Similarly to the Z string \cite{Vachaspati92}\cite{Achucarro00} in
the standard electroweak model, the pion string is not
topologically stable, since any field configuration can be
continuously deformed to the trivial vacuum in the QCD sigma
model. With finite temperature plasma, however one of us
(Nagasawa) and Brandenberger\cite{Nagasawa99} argued that the pion
string can be stabilized. They propose that the interaction of the
pion fields with the charged plasma generates a correction to the
effective potential and this correction reduces the vacuum
maniford $ S^3 $ of the zero temperature theory to a lower
dimensional sub-maniford $ S^1 $, which makes the pion string
stable. Moreover, it has been shown by numerical simulations that
semilocal strings, which are also not topologically stable, can be
produced at the phase
transition\cite{Achucarro00}\cite{Achucarro98}. In a similar way,
pion strings are expected to be produced during the QCD phase
transition in the early universe as well as in experiment of the
heavy ion collisions.. The strings will subsequently decay.

The pion string is a static configuration of the Lagrangian
${\mathcal{L}}_{\Phi}$ of Eq.(\ref{eq1}). In order to discuss the
pion string, we define new fields
\begin{eqnarray}
\phi =\frac{\sigma +i\pi ^0}{\sqrt{2}}
\end{eqnarray}
and
\begin{eqnarray}
 \pi ^{\pm }=\frac{\pi
^1\pm i\pi ^2}{ \sqrt{2}}.
\end{eqnarray}
The Lagrangian ${\mathcal{L}}_\Phi $ now can be rewritten as
\begin{equation}\label{eq2}
{\mathcal{L}}_\Phi =(\partial _\mu \phi ^{*})(\partial ^\mu \phi
)+(\partial _\mu \pi
^{+})(\partial ^\mu \pi ^{-})-\lambda (\phi ^{*}\phi +\pi ^{+}\pi ^{-}-\frac{%
f_\pi ^2}2)^2.
\end{equation}
For the static configuration, the energy functional corresponding
to the above Lagrangian is given by
\begin{equation}\label{eq3}
E=\int d^3x[\vec{\nabla} \phi ^{*}\vec{\nabla} \phi +\vec{\nabla}
\pi ^{+}\vec{\nabla} \pi ^{-}+\lambda (\phi ^{*}\phi +\pi ^{+}\pi
^{-}-\frac{f_\pi ^2}2)^2].
\end{equation}
The time-independent equations of motion are:
\begin{equation}
\nabla ^2\phi =2\lambda (\phi ^{*}\phi +\pi ^{+}\pi ^{-}-\frac{f_\pi ^2}2%
)\phi
\end{equation}
and
\begin{equation}
\nabla ^2\pi^+ =2\lambda (\phi ^{*}\phi +\pi ^{+}\pi
^{-}-\frac{f_\pi ^2}2)\pi^+.
\end{equation}
The pion string solution with a single winding number extremizing
the energy functional in Eq.(\ref{eq3}) is given by\cite{Zhang98}
\begin{equation}\label{eq4}
\phi =\frac{f_\pi }{\sqrt{2}}[1-\exp (-\mu r)]\exp (i\theta )
\end{equation}
and
\begin{equation}\label{eq5}
\pi ^{\pm }=0,
\end{equation}
here the coordinates $r$ and $\theta$ are polar coordinates in
$x-y$ plane (the string is assumed to lie along the $z$ axis),
$\mu^2=\lambda \frac{89}{144}f_{\pi}^2$, the energy per unit
length of the string is
\begin{equation}
E=[0.75+\log (\mu R)]\pi f_\pi ^2,
\end{equation}
where $R$ is introduced as a cutoff since for global symmetry the
energy density of the string solution is logarithmically
divergent. Generally $R$ is given by the horizon size or the
typical separation length between strings. The typical distance
between strings can be determined by the string number density at
the formation which is based on the the Kibble-Zurek mechanism and
the following evolution which is ruled out by the string tension
and the interaction between the string and the surrounding matter.
Thus the interaction between the strings would not be so
significant. In the following numerical calculation, we take $R=O(
{\rm fm})$, for other parameters we have $\lambda=9.877$,
$f_{\pi}=90$MeV, $m_{\pi}=140$MeV and
$m_{\sigma}=400$MeV\cite{Roder03}.

\section{The formation of the pion strings at LHC}
The order of the QCD chiral phase transition seems to depend on
the mass of the non-strange $u$ and $d$ quarks, $m_u\approx m_d$,
and the mass of the strange quark $m_s$. At the phase transition
temperature on the order of $150$MeV, heavier quark flavors do not
play an essential role. In the chiral limit, one can use
universality arguments to determine the order of the phase
transition. According to universality, the order of the chiral
transition in QCD is identical to that in a theory with the same
chiral symmetries as QCD, for instance, the $U(N_f)_L\times
U(N_f)_R$ linear sigma model for $N_f$ massless quark flavors.
This argument was employed by Pisarski and
Wilczek\cite{Pisarski84} who showed that for $N_f=2$ flavors of
massless quarks, the transition can be of second order, if the
$U(1)_A$ symmetry is explicitly broken by instantons, while for
three or more massless flavors, the phase transition for the
restoration of the $SU(N_f)_R\times SU(N_f)_L$ is first order.

So far, the formation of topological defects has been studied in
liquid crystal and superfluid experiments\cite{Kibble02}, which
are systems with global symmetry. It is generally believed that
the theoretical scenario that can be applied to determine the
defects initially formed immediately after second-order symmetry
breaking phase transition in this case is the Kibble-Zurek
mechanism. The basic picture of this mechanism is the following.
For a second phase transition, after the phase transition, the
physical space develops a domain like structure, with the typical
size of the domain being of the order of a relevant correlation
length $\xi$ (which depends on the nature of the dynamics of the
phase transition). Inside a given domain, the broken phase is
roughly uniform, but varies randomly from one domain to the other.
The string defects are to be formed in the junction of three or
more domains.

It is customary then for us to apply the Kibble-Zurek mechanism to
make an estimate of the density of pion strings during Pb-Pb
central collisions at the LHC with energy $\sqrt{s}=5.5 TeV$. At
the initial stage of the collision there exists manifestly partons
with very large cross section for gluon scattering, so the gluons
will reach equilibrium quickly with an initial temperature at
about $T_i=600$MeV corresponding to the time
$t_i=0.2$fm\cite{Heinz01}. If the entropy of the system is
conserved throughout the expansion, using the Bjorken model we
have that the thermal freeze out of the fireball occurs at
$t_f=25$fm when the temperature reaches $T_f=120$MeV. The
calculation in Ref.\cite{Braun99} shows that the quark-gluon
plasma can be formed over very large space-time volumes at the LHC
Pb - Pb collisions. The hydrodynamic model predicts that the
volume of such a plasma region evolves
as\cite{Bjork84}\cite{Hung98}\cite{Zhuan00},
\begin{equation}\label{volume}
V(t_f)=V(t_c)\frac{t_f}{t_c},
\end{equation}
where $V(t_f)=2\times 10^4$fm$^3$, while evolution of the
temperature is given by
\begin{equation}\label{temperature}
T(t)=T_i\left(\frac{t_i}t\right)^{\frac 13}.
\end{equation}
From Eqs.(\ref{volume}) and (\ref{temperature}) we obtain that the
time when the phase transition occurs at $T_c=170$MeV is
$t_c\simeq 8.793$fm, and the volume at the freezing out is
$V(t_f)=2 \times 10^4$fm$^3$.

When the LHC Pb - Pb collisions take place, a big fireball is
formed in the central region of the collision with the initial
temperature around $T(t_i)=600$MeV at the time $t_i=0.2$fm. The
fireball quickly reaches to the equilibrium state and it expands
rapidly with the volume and temperature given in
Eqs.(\ref{volume}) and (\ref{temperature}). During the period when
the temperature is higher than $T_c=170$MeV (before $t_c\simeq
8.793$fm), the fireball is in the QGP phase where the chiral
symmetry is unbroken. When the temperature of the fireball
decreases down to the critical point $T_c(t_c)=170$MeV and its
volume is about $V(t_c)\approx 7\times 10^3$fm$^3$, the fireball
undergoes a rapid second order chiral phase transition. At this
time the system is in an out of equilibrium dynamical state, and
the phase with broken chiral symmetry starts to appear due to the
fluctuations of the order parameter simultaneously and
independently in many separate regions of the expanding fireball.
Subsequently during the process with further cooling, these
regions grow and merge with each other to realize the new phase
with the broken symmetry all over the fireball. At the boundaries
where causally disconnected different regions meet, the order
parameter field does not necessarily match and a domain structure
is formed. This is essentially similar to the process of the
defect formation during the cosmological phase transitions in the
early universe.

As described by the Kibble-Zurek mechanism, the transition speed
can be given by the quench time $\tau _Q$
\cite{Zurek85}\cite{Volovik03}\cite{Kibble76},
\begin{equation}\label{q}
\tau _Q=\frac{T_c}{\mid dT/dt\mid _{t=t_c}}=3t_c.
\end{equation}
From the Ginzburg-Landau theory for the second order phase
transition, the quench time $\tau_Q $ can be deduced by the order
parameter relaxation time $\tau$ with a general form
\begin{equation}\label{tau}
\tau(T) =\tau _0\left(1-\frac{T}{T_c}\right)^{-1},
\end{equation}
where $\tau_0\sim \xi_0$ and $\xi_0$ is the zero temperature
limiting value of the temperature dependent coherence length
$\xi(T)$, which in this paper we take $\xi _0=1/m_\sigma\simeq
0.49$fm. When the temperature $T$ is close to $T_c$, we have
\begin{equation}\label{xi}
\xi(T)=\xi_0\left(1-\frac{T}{T_c}\right)^{-\frac{1}{2}}.
\end{equation}
As the temperature is below $T_c$ the order parameter coherence
spreads out with the velocity
\begin{equation}
c(T)\sim \frac{\xi}{\tau}=\frac{\xi_0}{\tau_0}
\left(1-\frac{T}{T_c}\right)^{\frac{1}{2}}.
\end{equation}
The pion strings are expected to be produced at the Zurek
freeze-out time $t_z$ when the causally disconnected regions have
grown together and the coherence is established in the whole
volume. At the Zurek freeze-out temperature $T(t_z)<T_c$, the
causal horizon is given by
\begin{equation}\label{xih}
\xi_{H}(t_z)=\int_{0}^{t_z}c(T)dt=\frac{\xi_0\tau_Q}{\tau_0}\left(1-\frac{T_z}{T_c}\right)^{\frac{3}{2}}.
\end{equation}
The causal horizon has to be equal to the coherence length
$\xi(t_z)$, then from Eqs.(\ref{tau})(\ref{xi})(\ref{xih}) we
obtain
\begin{equation}
t_z=t_c+\tau (t_z)=t_c+\sqrt{\tau _0\tau _Q}\approx 12.4fm ,
\end{equation}
and
\begin{eqnarray}
\xi _z=\xi _0(\tau _Q/\tau _0)^{1/4}\simeq \tau _Q^{1/4}\tau
_0{}^{3/4}\simeq 1.33fm.
\end{eqnarray}
Due to the Kibble-Zurek mechanism, a network of pion strings is
formed at the Zurek freeze-out time $t_z$ with a typical curvature
radius and separation of $\xi_z$. After that time, the network of
the pion string is going to evolve with the fireball expansions
until to the freeze-out time $t_f$.

\section{The evolution and decay of the pion string}
Similarly to the evolution of the cosmic string in the early
universe\cite{Brandenberger99}\cite{Brandenberger94}\cite{Vachaspati84},
when the temperature of the fireball falls down from the Zurek
temperature, $T_z\equiv T_i(\frac{t_i}{t_z}
)^{\frac{1}{3}}=151.6$MeV to $T_f$, the evolution of the string
network would obey the following procedure. Initially, at the
Zurek time $t_z$, the pion string has a typical curvature radius
and separation of the correlation length $\xi_z$, which then
increases rapidly and eventually approaches a scaling solution in
which $\xi(t)\sim t^a$. In the case of the pion string in the
heavy ion collision experiments, since the volume of fire ball
obeys the following law $V(t)\sim t$, we make the assumption that
$a=\frac{1}{3}$ for our case from time $t_z$ up to $t_f$. Hence,
the correlation length of the pion string at time $t_f$ is
\begin{eqnarray}
\xi_f=\xi(t_f)=\xi_z\left(
\frac{t_f}{t_z}\right)^{\frac{1}{3}}\simeq 1.69fm .
\end{eqnarray}
The pion string ceases to evolve at the time $t_f$ and decays
because of the fireball disappearance. Hence we have to find out
the size and number of loops at the time $t_f$.

Note that in our situation, closed string loops have a dominant
contribution to the total energy of the string. This is because at
the freezing out, the string evolution is still ruled out by the
frictional force by the surrounding matter so that the free motion
of the string is not realized. In addition, the expansion of the
system is too rapid that the initial Brownian string distribution
will be conserved. Then the initial structure of the string
network partially remain and the spatial trajectories of strings
are very much complicated. Thus we take the initial pion string
network as that of the Brownian one and regard that the
distribution of these loops does not change with time, the size of
the loops are conformally stretched during the expansion of
fireball and a simple scaling can be realized. By using the scale
invariance of Brownian string described by Vachaspati and
Vilenkin\cite{Vachaspati84}, we get the distribution of number
density of the pion string with the length between $l$ and $l+dl$
at the freeze out temperature, $T_f$,
\begin{equation}\label{eq01}
dn(l)=K\xi_{f}^{-\frac{3}{2}}l^{-\frac{5}{2}}dl,
\end{equation}
where the parameter $K$ is approximately in the range of $0.01\sim
0.1$\cite{Volovik03}.

Integration of Eq.(21) over $d l$ will result in the total number
of the string loops. Note that the string width $r_0\sim
\frac{1}{\mu}$ gives a minimum length of the pion string at time
$t_f$, $l_{0}=2 \pi r_0\simeq 5.6$fm and the longest string is
also constrained by the volume of the system. From Eq.(\ref{eq01})
we have the total number of pion strings
\begin{eqnarray}\label{eq03}
N(0)=\frac{2 K V(t_f)}{3 \xi_f^{\frac{3}{2}}
l_{0}^{\frac{3}{2}}}=459K.
\end{eqnarray}
Since $K$ varies from 0.01 to 0.1 the total number of pion strings
varies between $N(0)\simeq 4$ and $N(0)\simeq 45$.

In the immediate aftermath of the phase transition, when the
temperature is still close to $T_c$, the string tension remains
small and motion of the strings is heavily damped by the
frictional effects of the surrounding high-density medium. The
mechanism of Nagasawa and Brandenberger implies that pion strings
might effectively be stable in this high-density medium until the
thermal freeze out time $t_f$. After that, the string tension
approaches its zero-temperature value and the motion of the
strings is effectively decoupled from the surrounding medium,
there will not be corrections to the effective potential from the
thermal bath and pion strings undergo the second phase transition.
Even though pion strings undergo the second phase transition under
the temperature $T_f$, pion strings will not decay immediately and
can still survive sometime below the freeze out time $t_f$, since
the pion strings undergo a core phase transition and lose their
central structure where the field strength equals to zero but
still preserve the winding number of neutral
components\cite{Nagasawa03}. For simplicity we assume that pion
strings can survive after the decoupling time and all pions which
are eventually emitted from pion strings will be completely
incoherent with the rest of pions.According to the work of
Ref.\cite{Zhuan00}, the produced pions in the interval $T>T_f$
have time to be thermalized before the freeze out time, they lead
to an enhancement of thermal pions, while in the interval $T<T_f$
the produced pions do not get chance to be thermalized, they
result in a nonthermal enhancement of pions with low momentum.
Thus it is expected that the resultant pion spectrum with the pion
string will contain the nonthermal pions which can be taken as the
evidence of the pion string.

All pion strings will decay into the sigma particles and neutral
pions. To estimate the numbers of the particle produced we notice
that for the ansatz Eqs.(\ref{eq4}) and (\ref{eq5}), the sigma
field in Eq.(\ref{eq3}) contributes about $50\%$ of the total
energy of the string. Due to energy conservation half of the
string energy should convert into that carried by the sigma
particles. The remaining $50\%$ of the string energy will go to
the neutral pions. For global string such as the axion string
 \cite{Yamaguchi99} one expects the mesons produced from the decay of
the pion strings with length $l$ have a typical momentum $p\sim
1/l$. Using Eq.(\ref{eq01}), we obtain that the total number of
sigma particle $N_{\sigma}$ emitted from pion strings within a
fireball is about $100$(K=0.1), $43$(K=0.05), $22$(K=0.03) and
$5$(K=0.01), respectively. And the total number of neutral pions
$N_{\pi^0}$ is about $332$(K=0.1), $148$(K=0.05), $80$(K=0.03) and
$19$(K=0.01). As mentioned above it is expected that the
eventually resultant pion spectrum will have a nonthermal
enhancement at low momentum region because all produced pions from
pion strings are distributed at low momentum.

It can be seen that the most of the sigma particles and the
neutral pions have a relatively low momentum and these particles
are nonthermal particles. The momentum distribution of the sigma
from the pion string decay is given by
\begin{eqnarray}
\frac{dN_{\sigma}(p)}{dp}=\frac{KV(t_f)\sqrt{p}}{D_1\xi_f^{\frac{3}{2}}},
\end{eqnarray}
where the normalization factor $D_1=0.476$. The momentum
distribution of the neutral pions is given by
\begin{eqnarray}
\frac{dN_{\pi^0}(p)}{dp}=\frac{KV(t_f)\sqrt{p}}{D_2\xi_f^{\frac{3}{2}}},
\end{eqnarray}
where the normalization factor $D_2=0.142$. The averaged momentum
of sigma and neutral pion is $\langle p\rangle \simeq 21.1$MeV.
The neutral pions and sigma particles are dominant in the low
momentum region, and the momentum distribution of the pions
produced at the decay of the pion string can be taken as a
distinctive signal of the formation of the pion string in heavy
ion collisions.

\begin{figure}[htbp]
\includegraphics[scale=0.8]{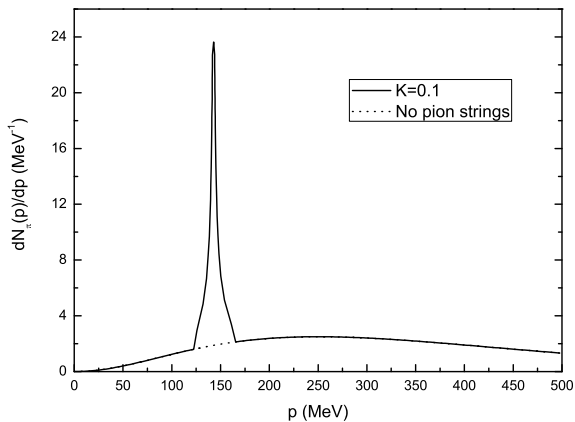}
\caption{The momentum spectra of the total pions from the pion
string and thermal pions by taking parameter K=0.1 and without
considering the large decay width of the sigma. The indirect pions
($\pi^0$'s, $\pi^\pm$) via $\sigma$ decay from the pion strings
are mostly distributed around $\langle p \rangle\sim 143.2 MeV$.}
\label{fig1}
\end{figure}

The sigma particles from pion strings will decay equally into
neutral and charged pion mesons\footnote{An early study on the
effects of pion string in heavy ion collision has been performed
in Ref.\cite{Balachandran02}, but they have not quantitatively
estimated the numbers of the pions produced from the pion string
decay.}. As we know, the $\sigma$ particle has a large decay width
and it is more complicated issue to get the real distribution of
the pions from the $\sigma$ meson. In order to get the simplified
results, we ignore its large decay width and take it as a stable
particle. Since the momentum of the sigma can be approximately
neglected, for $m_\sigma = 400$MeV these pions from the sigma
decay will have momentum around $143$MeV. In Fig.\ref{fig1} we
plot the distribution of these pions as a function of the momentum
together with the thermal pions calculated in Ref.\cite{Zhuan00}.
Numerically there are about $200$ nonthermal pions mostly
distributed at $p\sim 143$MeV for $K=0.1$(In the following
discussion, for simplicity, we always take the parameter $K=0.1$
for the case of the sigma decay.).

\begin{figure}[htbp]
\includegraphics[scale=0.8]{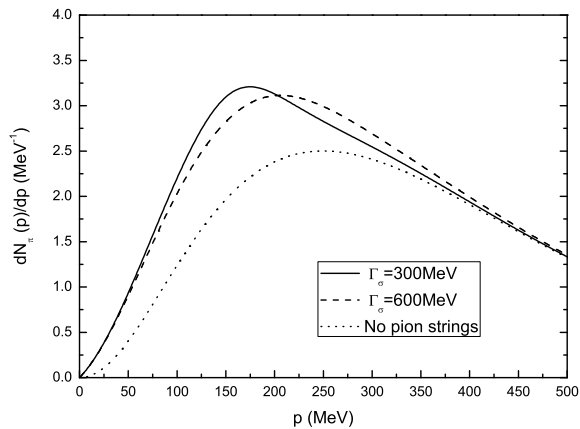}
\caption{The momentum spectra of the total pions from the pion
string and thermal pions by taking parameters
$\Gamma_{\sigma}=300MeV$ and $\Gamma_{\sigma}=600MeV$ and $K=0.1$.
The indirect pions ($\pi^0$'s, $\pi^\pm$) via $\sigma$ decay from
the pion strings are broadly distributed from $p\sim 0 MeV$ to
$p\sim 500MeV$.} \label{fig2}
\end{figure}

In the above discussion, we get the simplified results by ignoring
the large decay width of the sigma and take it as a stable
particle. In fact, the $\sigma$ particle is a broad resonance and
have a decay width, then the narrow peak in Fig.\ref{fig1} will
definitely smoothen out. In order to get a more reality figures,
we should take into account the decay width of sigma and redraw
the figure. Mass and width of $\sigma$ particle are somewhat model
dependent, in this paper, we take the Breit-Wigner function
as\cite{Jing-Zhi04}
\begin{eqnarray}\label{eq6}
BW_{\sigma}=\frac{1}{m_{\sigma}^2-s-im_{\sigma}\Gamma_{\sigma}},
\end{eqnarray}
where $\Gamma_{\sigma}$ is a constant. From Eqs.(26)(28) and
(\ref{eq6}), we plot the distribution of these pions together with
the thermal pions as a function of the momentum and the decay
width of sigma by taking the parameters as
$\Gamma_{\sigma}=300MeV$ and $\Gamma_{\sigma}=600MeV$ and $K=0.1$
in Fig.\ref{fig2}. The averaged momentum of indirect pions via
sigma decay is $\langle p\rangle \sim 152.5$MeV and $\langle
p\rangle \sim 175.9$MeV according to $\Gamma_{\sigma}=300MeV$ and
$\Gamma_{\sigma}=600MeV$ respectively. Numerically there are also
about $200$ nonthermal pions broadly distributed at the momentum
region $p\sim 0-500$MeV for $\Gamma_{\sigma}=300MeV$ and
$\Gamma_{\sigma}=600MeV$. From the Fig.\ref{fig2}, the narrow peak
in Fig.\ref{fig1} definitely smoothens out and nearly disappears.
These nonthermal pions can hardly be distinguished from the
thermal pions, then one of the possible signal of pion stings is
not significant because of the large decay width of sigma.

\begin{figure}[htbp]
\includegraphics*[scale=0.8]{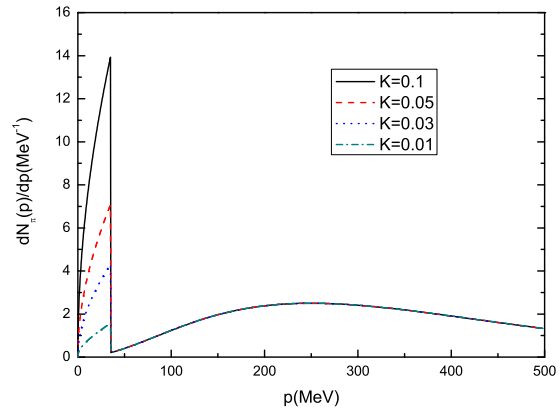}
\caption{The momentum spectra of the nonthermal neutral pions
($\pi^0$'s) emitted from the pion strings together with the
thermal pions. The nonthermal pions are mainly distributed around
$p \sim 0-35MeV$.} \label{fig3}
\end{figure}

Since the string configuration violates the isospin symmetry the
direct production of the pions from the pion string decay is only
for the neutral pion, not the charged pion \footnote{The charged
pion strings are expected to be also produced, however according
to the mechansim of Nagasawa and Brandenberger the charged pion
string will not be able to be stabilized in the plasma,
consequently they will decay away and their decay products will be
thermalized before the freezing out. }. In Fig.\ref{fig3} we plot
these neutral pion distribution together with the thermal pions.
Numerically there are about $N_{\pi^0} \sim 20-300$ nonthermal
pions distributed in the low momentum region with $\langle
p\rangle \sim 21$MeV. From the Fig.\ref{fig3}, all the neutral
pions from pion strings are distributed at the low momentum and
the ratio of neutral to charged pions from pion strings violates
the isospin symmetry, this can be taken as the possible signal of
pion strings.

\section{Summary and discussion}
We have investigated the effects of the pion string in the
experiment of the heavy ion collisions. Following the Kibble-Zurek
mechanism pion strings are expected to be formed in LHC Pb - Pb
collision at energy $\sqrt{s}=5.5$ TeV, then decay after the
freezing out time into pions. These pions are mostly distributed
in two separated low momentum regimes. These effects are expected
to be observable and differ from predictions of other
models\cite{Zhuan00}\cite{Stephanov99}. The pion enhancement in a
small window around the nonthermal momentum $p_0\simeq 21.1$MeV
for neutral pions. While for other nonthermal pions, the situation
is completely different. If we ignore the large decay width of the
sigma and take the sigma as a stable particle, there are the pion
enhancement in a small window around the nonthermal momentum
$p_0\simeq 143$MeV, but actually the sigma has the large decay
width, then the peak will smoothen out and almost tend to
disappear. So it is difficult to detect this part of nonthermal
pions in experiment. All the resultant pion spectrum depend
strongly on how long the pion string can survive below the freeze
out time $t_f$.

In this paper, we have made the assumption that pion strings can
survive after the decoupling time, then both the neutral pions and
sigma particles emitted from pion strings do not get chance to be
thermalized. On the other hand, we do not exclude the situation in
which part (even all) of pion strings will decay into pions and
sigma particles when the time is very close to the decoupling
time, and such produced pions from pion strings will be
thermalized by the final state interactions and the peak in the
pion spectra due to the pion string decay will disappear partly
(or completely). However, even though this situation is happened,
there still have possible signals of the pion string produced.
Then pion string decay can lead to experimentally observable
anomlies which are very similar to the DCC(Disoriented Chiral
Condensate) decay. It is the ratio of neutral to charged pions,
$r=\frac{n_0}{n_0+n_{ch}}$, here $n_0$ is the number of neutral
pions while $n_{ch}$ is corresponding to charged pions, which is
different from what is naively expected ($\frac{1}{3}$) if there
are pion strings produced during chiral phase transition. Also
enhancement of sigma which in turn decay into pions around the
decoupling time will also result in nonthermal pions, though thses
nonthermal pions is difficult to be detected because of the large
decay width of the sigma.

Therefore, in order to obtain more reliable conclusions, we need
to know more detail of the process of the pion string decay, the
sigma decay and behaviors of the fire-ball at freeze out time in
heavy ion collisions.

\section{Acknowledgments} The authors wish to thank Nicholas
Petropoulos, Xinnian Wang and Ning Wu for useful discussions and
correspondence. We also thank Xinggang Wu for discussions and help
in the numerical calculation. M.N. is grateful to IHEP for kind
hospitality. This work is supported in part by the National
Natural Science Foundation of People's Republic of China. The
visit of M.N. to IHEP is financially supported by the China-Japan
exchange program.

\end{document}